# BUILDING AN AUTOMATED GESTURE IMITATION GAME FOR TEENAGERS WITH ASD

*Linda Nanan VALLÉE[1*], Christophe LOHR[2], Sao Mai NGUYEN[2], Ioannis KANELLOS[2], Olivier ASSEU[1]*

[1]*Ecole Supérieure Africaine des TIC, Abidjan, Côte d'Ivoire*
[2]*Institut Mines Telecom Atlantique de Bretagne, Lab-STICC, France*
*linda.vallee@esatic.edu.ci

*Abstract*

Autism spectrum disorder is a neurodevelopmental condition that includes issues with communication and social interactions.

People with ASD also often have restricted interests and repetitive behaviors.

In this paper we build preliminary bricks of an automated gesture imitation game that will aim at improving social interactions with teenagers with ASD. The structure of the game is presented, as well as support tools and methods for skeleton detection and imitation learning. The game shall later be implemented using an interactive robot.

**Keywords:** Autism, Imitation, Artificial intelligence, Human-machine interaction, Gesture recognition

## I. INTRODUCTION

Information and communication technologies have contributed to the social and cognitive stimulation of children with autism spectrum disorder (ASD). This seems to be partly because children with ASD are more comfortable with predictive and repetitive behaviours [1], which can be implemented through algorithms. In particular, artificial intelligence algorithms are used in various fields, from language or gesture recognition to image classification [2]. These functions are convenient for a system to interact with a human being.

Furthermore, robots seem to prove useful with autistic children because of simpler face expressions than those of human beings [3]. Robots can perform gesture imitation learning [4].

For a gesture to be recognized, the human body must first be correctly represented.

Joint angles and joint positions can be used to represent human motion in spaces more suitable than the Euclidian one. Human postures and motion are nonlinear therefore sometimes represented in the Riemannian space [5]. An example of such body representation lies in [6].

The system learns body gesture by observation of several demonstrations.

Then, probabilistic models, such as Hidden Markov Models or Gaussian mixture models (GMM), can help determine if imitation attempts match a target movement.

The work performed in [7, 8] shows how the skeletal model, the representation of the human body in the Riemannian space as well as GMM were used for automated physical rehabilitation exercises. Within our context, skeleton detection and gesture recognition methods will be used in an attempt to improve the participants' abilities to interact with others.

The imitation process is a pillar for learning and interacting. [9] demonstrates that autistic children are able to imitate.

Through imitation practice it appears possible to improve imitation abilities and even reduce the degree of autism [10]. The experimentation consisted of human caregivers performing several simple imitation games with the children. In order to evaluate the quality of the imitation, the scale defined by Nadel was used.

Although this work is interesting, it does not build on modern tools, such as imitation learning algorithms or robots.

In [11, 12] however, it is shown how an interactive robot could bring positive results in terms of social interactions with autistic children.

In this paper, we redirect the idea of a gesture imitation learning algorithm towards the improvement of social interactions with autistic teenagers and preteens. As steps on the path to this purpose, skeleton detection and imitation learning methods are described.

The following section of this paper describes the methodology before simulation results for skeleton detection are presented in section III.

Section IV contains a short summary of the work as well as some perspectives.

## II. METHODS

Work sessions are regularly held with autism professionals, autistic teenagers and their parents, with the aim of keeping connected with real-life challenges and needs.

This is crucial in order to set up an adequate environment and propose a game that will actually be usable.

The test environment shall be quiet and free of distractions.

In order to ease interactions with the participants, autism professionals recommended that our gesture imitation game would start with greetings and pairing stages, followed by several imitation modules: for instance, one induced, one spontaneous, with or without objects.

A skeleton detection method is needed from the beginning and all through the game. The Tensorflow implementation of the Openpose algorithm was tested. Tensorflow is a library used to train and execute neural networks for element classification like in gesture recognition.

The Openpose algorithm allows for the initial detection, through the computer camera, of the skeleton of the participant, then successive poses can be recognized.

Figure 1 displays a schematic representation of the human body as used in [13].

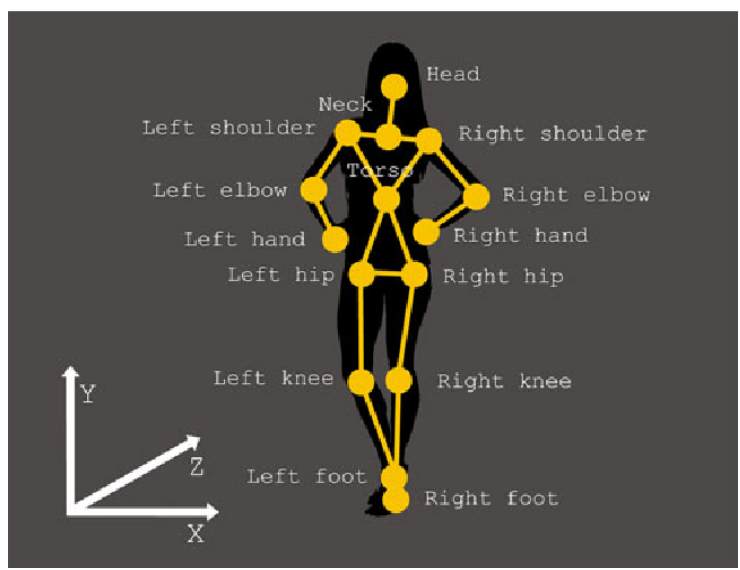

**Figure 1:** schematic human body representation

In order to detect a pose, the orientation and position of relevant joints are to be considered. Joint positions are not absolute but normalized relative positions, computed from their absolute positions relatively to the absolute position of the spine shoulder. Their normalization is done using the spine bone length.

The 3D Riemannian manifold as described in [5] is more suitable to represent human joint orientations because it is nonlinear. Local tangent spaces must be considered to compute standard statistics.

Papers [6, 8] describe the detailed approach for computing the mean μ, also called the Riemannian center of mass, of N points $p_i$ on the human pose space then the covariance matrix, allowing for the learning of a Gaussian Mixture Model:

$$p(x) = \sum_{k=1}^{K} \phi_k \, N(x \; \mu_k, \Sigma_k) \qquad (1)$$

where x encodes both the human pose $y_t$ and the timestamps t, K is the number of Gaussians, $\phi_k$ the weight of the k-th Gaussian, $\mu_k$ the Riemannian center of mass of the k-th Gaussian computed on the manifold and $\Sigma_k$ the covariance matrix of the k-th Gaussian. The parameters $\phi_k, \mu_k$ and $\Sigma_k$ are learned using Expectation-Maximization on the human pose space.

It is once a model is learned that an optimal sequence can be generated, approximating the sequence of poses with a single Gaussian:

$$p(\hat{x} \; t) \approx N(\hat{\mu}, \hat{\Sigma}) \qquad (2)$$

This is performed using Gaussian Mixture Regression.
It is at the tangent space at $\hat{\mu}$ that $\hat{\Sigma}$ is computed.
Participants' motion sequences will be compared to the ideal movement, achieved by evaluating $\hat{x}$ along time.

In [8] within the framework of physical rehabilitation, the following formula is used in order to estimate the probability that a given sequence X was generated by the learned GMM:

$$ln\,(p(X \; \phi, \mu, \sigma) = \sum_{t=1}^{T} \ln \left( \sum_{k=1}^{K} \phi_k N(x_t \; \mu_k, \Sigma_k) \right) \qquad (3)$$

Then temporal segment and body part analysis are performed to determine which section of the exercise is not performed correctly as well as which specific body part is responsible for the error.

In our context, this additional processing will not be necessary as the gesture imitation by the teenager with ASD does not have to be perfect, and it is not relevant to explain the detected errors. What is of interest rather, is that there is an actual imitation attempt and a social relationship is established between the teenager with ASD, the human caregiver and later on the robot.

## IV. SIMULATION RESULTS AND DISCUSSION

For human pose detection, Tensorflow Pose Estimator was implemented following standard steps.

We executed *run_webcam.py* with the model and format of our choice, and obtained the following results captured through the camera **(figures 2 and 3)**.

On the first capture (fig. 2a), our participant stands from side-on with the left knee raised.

On the second capture (fig. 2b), our participant appears from the front, with hands on the waist.

In both cases, all body parts are correctly detected.

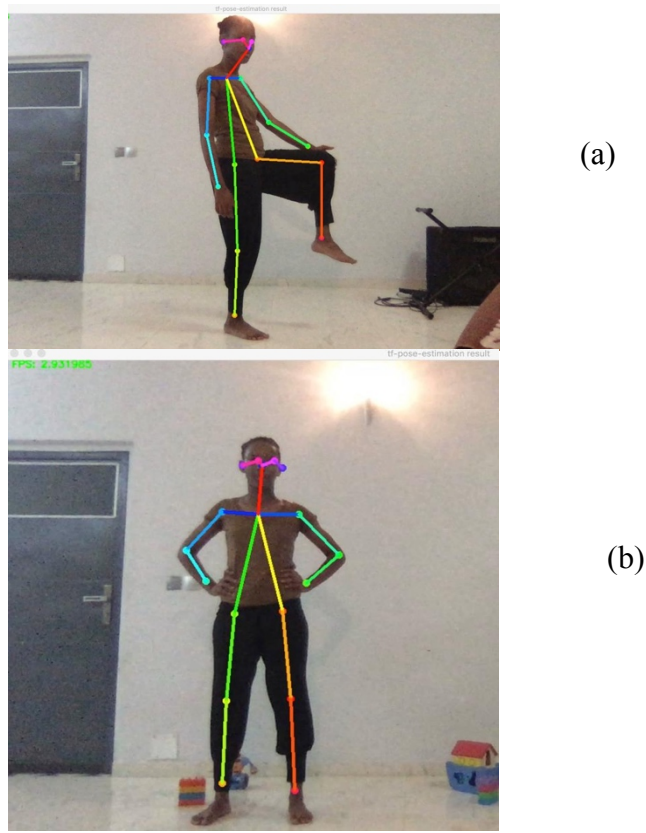

(a)

(b)

**Figure 2:** successful experimental results for skeleton detection

On figure 3, the first picture (fig. 3a) shows two participants from the front, the shortest one (toddler) before the other (teenager).

The teenager's legs are not correctly detected as there is body occlusion.

(Fig. 3b) is a three-quarter photograph of the teenager participant bending on her knees, with hands raised in front of her.

The left arm is not detected. Neither is the lower part of the leg lying on the floor.

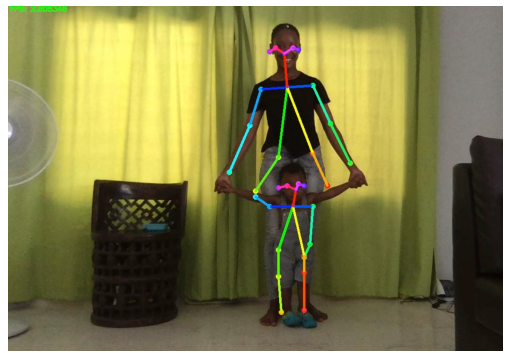

(a)

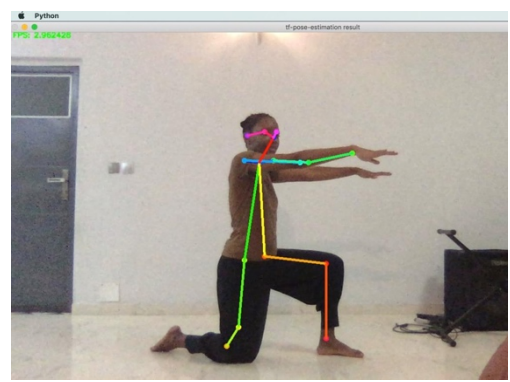

(b)

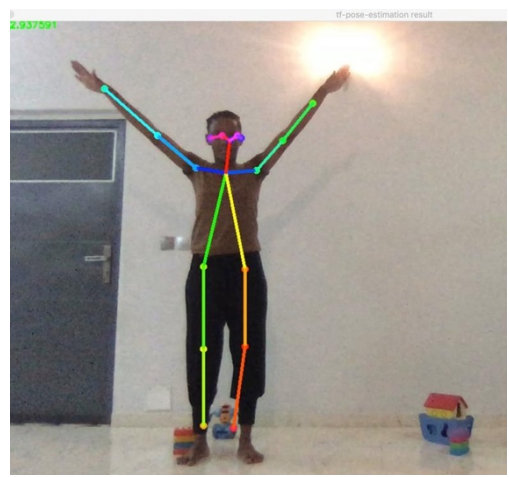

(c)

**Figure 3:** unsuccessful experimental results for skeleton detection

On the third capture (fig. 3c), the participant is seen from the front with arms raised in a cross-like posture.

All body parts are correctly detected, apart from the left ankle that is positioned by the algorithm on the small robot toy behind it. The robot located just behind the ankles may be considered as noise in the detection process.

Percentage of success can be represented as the number of detected body parts over the total number of body parts, which allows us to build **Table 1**:

| Figure | # detected body parts (/14) | Percentage of success |
|---|---|---|
| 2a | 14 | 100 % |
| 2b | 14 | 100 % |
| 3a - teenager | 10 | 71% |
| 3a - toddler | 11 | 79% |
| 3b | 9 | 64% |
| 3c | 13 | 93% |

In a real-life setting, skeleton detection can be performed using the Openpose algorithm. However, the environment settings, as well as the body postures included within the game, must be carefully chosen in order to make sure that those are correctly detected.

For instance, lighting should be sufficient and body occlusion should be limited.

Formula (1), (2) and (3) are then used for gesture learning and recognition.

**V. CONCLUSION**

In this work, after reviewing previous studies, several aspects of our methodology were presented: imitation game constraints and structure, technical methods for body detection, representation and gesture recognition.

Skeleton detection results in a real-life environment were then presented.

The complete serious game will be developed with the aim of improving social interactions with autistic teenagers.

Implementing the game through an interactive robot will be an interesting perspective as robots display simplified face expressions, which autistic people can better handle than human face expressions.

Vallée, L. N., Lohr, C., Nguyen, S. M., Kanellos, I., and Asseu, O. (2019). BUILDING AN AUTOMATED GESTURE IMITATION GAME FOR TEENAGERS WITH ASD, Far East Journal of Electronics and Communications, 22, (1-2), 19--28. http://dx.doi.org/10.17654/EC023010001## ACKNOWLEDGEMENT

Here is the opportunity to express deep gratitude to Maxime Devanne for his inspiring research work, as well as to Monique and William for their active participation in the skeleton detection tests.